\documentclass[11pt]{article}
\usepackage{moriond,epsfig}
\usepackage[latin1]{inputenc}
\usepackage[OT1]{fontenc}
\usepackage{wrapfig}
\usepackage{amssymb}

\def\Journal#1#2#3#4{{#1} {\bf #2}, #3 (#4)}

\def\NIMA{{\em Nucl. Instrum. Methods} A}
\def\MPLA{{\em Mod. Phys. Lett.} A}
\def\NPA{{\em Nucl. Phys.} A}

\def\NPBS{{\em Nucl. Phys.} B {\em (Proc. Supp.)}}

\newcommand{\celcius}{\,^{\circ}\mathrm{C}}

\def\be{\begin{equation}}
\def\ee{\end{equation}}
\def\bea{\begin{eqnarray}}
\def\eea{\end{eqnarray}}

\begin{document}
\vspace*{4cm}

\title{A SEARCH FOR NEUTRINOLESS DOUBLE BETA DECAY: FROM NEMO-3 TO SUPERNEMO}

\author{ Yu.A. SHITOV on behalf of the SuperNEMO Collaboration}

\address{Physics Department, Imperial College\\
Prince Consort Rd, London SW7 2BW, UK\\
$^*$E-mail: y.shitov@imperial.ac.uk}

\maketitle\abstracts{ The SuperNEMO project aims to search for
neutrinoless double beta decay ($0\nu\beta\beta$) up to a sensitivity
of 10$^{26}$ years for the $0\nu\beta\beta$ half-life (down to $\sim$ 50~meV
in the effective Majorana neutrino mass), using $\sim$100 kg of source
and a `tracko-calo' detector. The current status of the 2006--2010
R\&D programme is discussed here.}

\section{Introduction}

Discovery of neutrinoless double beta decay, $0\nu\beta\beta$:
$(A,Z) \rightarrow (A,Z+2) + 2e^-$, which is forbidden in the Standard Model
due to lepton number conservation, would prove the Majorana nature and
reveal new fundamental properties of the neutrino. The hierarchy and absolute
mass scale of the neutrino eigenstates would be determined in the case where
$0\nu\beta\beta$-decay is driven by light neutrino exchange, while new
physics could be tagged in the case of other possible
$0\nu\beta\beta$-decay mechanisms, such as
right-handed currents, R-parity violation of SUSY, etc~\cite{Suh05,Ver05}.

The principal difference in experimental techniques is whether the two
electrons emitted in the $\beta\beta$-decay are measured directly
(tracking + calorimetry or TPC) or not (geochemistry or calorimetry
only). Pure calorimeters (germanium semiconductors and
bolometers) are the $\beta\beta$-sources themselves and thus only
measure the total energy deposited by both electrons. 

In comparison with calorimeters, the direct methods currently have worse
efficiency and energy resolution, but better background rejection
and the possibility of measuring different isotopes. 
However, the most important feature is that the individual energies and
trajectories of both electrons can be measured. Obtaining this unique 
information is the only way to probe the decay mechanism once a
$0\nu\beta\beta$-signal has been found by any experiment.

The SuperNEMO project is the next step in direct experimental
$0\nu\beta\beta$-decay searches based on the `tracko-calo' technique
of the NEMO series of experiments, including the latest 
currently running NEMO-3 detector~\cite{NEMO3}.  
Inspired by the success of these experiments, the NEMO/ SuperNEMO 
Collaboration~\footnote{Includes $\sim$100
physicists from 12 countries (http://nemo.in2p3.fr).} has embarked on an
 R\&D programme (since 2006) to design a detector with sensitivity 
down to $\sim$ 50~meV in the effective Majorana neutrino mass 
(up to 10$^{26}$ years for the $0\nu\beta\beta$ half-life) from measurements 
of $\sim$100 kg of source.

The SuperNEMO basic features and the current status of key R\&D
studies are presented in this article.

\section{Status of SuperNEMO R\&D}

The SuperNEMO project will extrapolate the NEMO-3 `tracko-calo'
technology to the new scale with the principal parameters shown in
Table~\ref{parameters}.

\begin{table}[hbt]
\caption{Comparison of the main NEMO-3 and SuperNEMO parameters.}
\label{parameters}
\begin{tabular}{|l|c|c|} \hline
Parameter & NEMO-3 & SuperNEMO  \\ \hline
Isotope & $^{100}$Mo & $^{82}$Se or other \\ \hline
Mass, kg    & $7$ & $100$+ \\ \hline
Efficiency, \% & $18$ & $\simeq 30$ \\ \hline
Energy resolution at 1 MeV (3 MeV) e$^-$, FWHM in \% & $\sim 12$
($\sim 8$) & $\sim 7-8$ ($\sim 4$)\\ \hline
$^{208}$Tl in foil, $\mu$Bq/kg & $<20$ & $<2$ \\ \hline
$^{214}$Bi in foil, $\mu$Bq/kg & $<300$ & $<10$ (only for $^{82}$Se) \\ \hline
Internal background ($^{208}$Tl, $^{214}$Bi), counts/full mass/year & 0.5 & 0.5 \\ \hline
$T_{1/2}^{0\nu\beta\beta}$ sensitivity, $\cdot$10$^{26}$~years & $>0.02$ & $>1$ \\ \hline
$<m_\nu>$ sensitivity, meV& $300$--$900$ & $40$--$110$ \\ \hline
\end{tabular}
\end{table}

\subsection{Design}

\begin{figure}[hbt]
\begin{center}
\psfig{figure=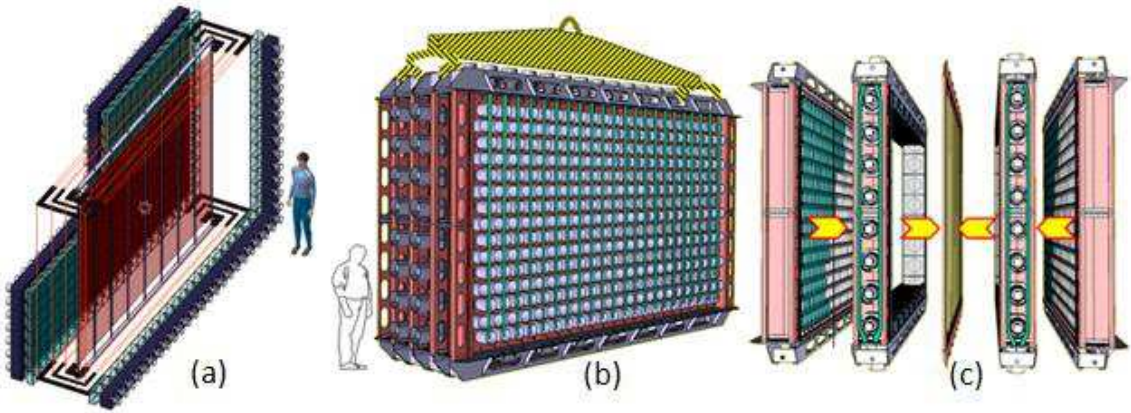,width=0.95\linewidth}
\caption{The principal SuperNEMO design (a) has a rectangular source foil sandwiched between two planar tracking chambers following the calorimeter walls; the demonstrator assembly (b) and its modular structure (c).}
\label{snemo_design_demo}
\end{center}
\end{figure}

The SuperNEMO detector (see Fig.~\ref{snemo_design_demo}) will follow a modular concept (20 units) with $\sim$5~kg of isotope per 5~$\times$~4~$\times$~1 m
module. 
Electrons emitted from a thin ($\sim$40 mg/cm$^2$)
$\beta\beta$-source foil in the middle of the module traverse a  
tracking chamber (2000-3000 wire drift cells operated in Geiger mode) 
before entering a calorimeter  ($\sim$ 600 channels: organic scintillator blocks
 coupled to PMTs). 

The alternative design ``bar mode'' (long scintillator bars instead of blocks) has been backed as the ``plan B'' option.

\subsection{Isotope Choice}
The physics criteria for the isotope choice are: i) large
$Q_{\beta\beta}$ to give a big phase space factor and better
background rejection, ii) large $T_{1/2}^{2\nu\beta\beta}$ to reduce
unavoidable $2\nu\beta\beta$-background, iii) large nuclear matrix
element (NME) to enhance the decay rate. Unfortunately, the latter is
rather unreliable as NME uncertainties remain quite large despite
recent progress in the development of calculation methods~\cite{Suh05}.

High natural isotope abundance, easy enrichment, radiopurification
and foil preparation are practical criteria required to produce 100~kg
of ultra-radiopure thin sources at a reasonable price. The main
candidate for SuperNEMO is $^{82}$Se. 

The full production chain is being studied now for selenium: i)
centrifugation enrichment has been tested~\footnote{Enrichment and
most of the purification have been funded by ILIAS
(http://www-ilias.cea.fr).}  producing 3.5~kg of $^{82}$Se; 100~kg
could be produced in 3 years; ii) purification by two methods
(chemical and distillation) has been carried out and checked at the kg scale;
iii) foil production has been redone with NEMO-3 technology and a new
technique is being tested.

The second option $^{150}$Nd could be a promising isotope for measurements (less background restrictions and possible physics benefits) but the possibility of its large scale enrichment is still unclear. 
Recently the SuperNEMO Collaboration has initiated R\&D studies in Russia with the aim of enriching $^{150}$Nd via hot gas ($\simeq$ 80$\celcius$) centrifugation, which looks promising for large scale isotope production.

\subsection{Calorimeter}

The energy resolution is a key factor in discriminating a 
$0\nu\beta\beta$-signal from $\sim$10$^5$--10$^6$ times as much unavoidable
$2\nu\beta\beta$-background. To reach a factor of two improvement
relative  to NEMO-3 (see Table~\ref{parameters}), with the
$\sim$1000~m$^2$ of detection surface in SuperNEMO, is a
challenging task as the technology has already been well tuned over
many years. The search for the best design  includes: 
i) tests of different scintillator materials (plastic, liquid,
non-organic) produced by improved technology where possible; 
ii) maximisation of light collection, choosing optimal scintillator shape
and size, new and improved reflector coating materials; 
iii) development of new ultra-low background, high quantum efficiency (HQE)
PMTs, working closely with the Hamamatsu, Photonis and ETL companies; 
iv) design of a technical implementation of the calorimeter. 

\begin{figure}[h]
\begin{center}
\psfig{figure=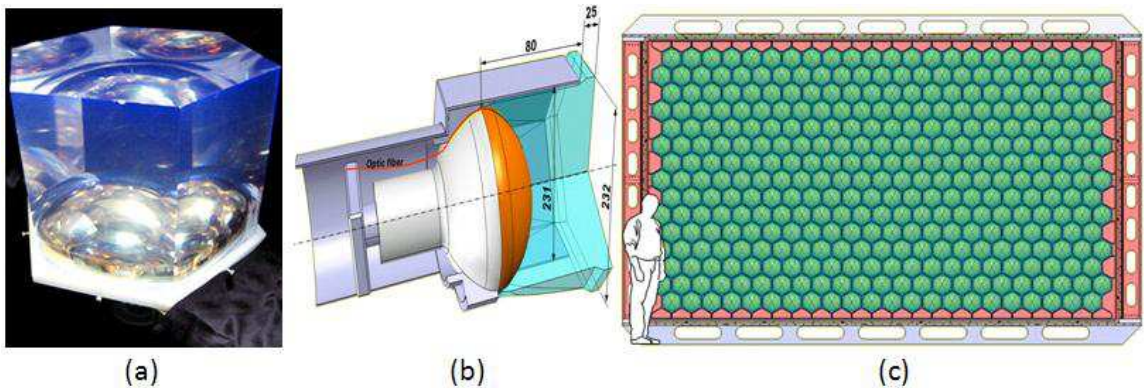,width=0.95\linewidth}
\caption{Developed calorimeter cell prototype with required resolution (a) as well as current design of calorimeter detection cell (b) and the whole wall (c).}
\label{snemo_calo}
\end{center}
\end{figure}

As a result of all tremendous efforts the required resolution (4\%@3 MeV) has been demonstrated with 28~cm hexagonal PVT blocks ($\geq$ 10~cm thick) directly coupled to 8-inch HQE Hamamatsu and Photonis PMTs (see  Fig.~\ref{snemo_calo}).

\subsection{Source radiopurity}

The most of dangerous internal sources are $^{208}$Tl and $^{214}$Bi
contaminations which must be reduced by factors 10 and 30, respectively, in
comparison with the NEMO-3 detector (see Table~\ref{parameters}).  Such
ultra-radiopurity is beyond the sensitivity of standard
low-background measurement techniques (1 kg~$\times$~month exposure
in $\sim$400~cm$^3$ HPGe). 

\begin{figure}[hbt]
\begin{center}
\psfig{figure=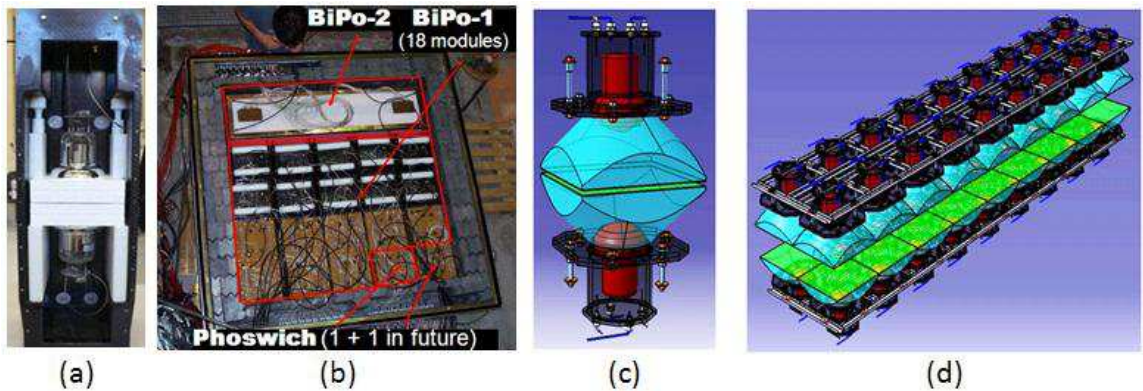,width=0.95\linewidth}
\caption{BiPo-1 capsule (a) and BiPo host site in LSM (b), capsule (c) and module (d) designs of BiPo-III detector currently under construction.}
\label{snemo_bipo}
\end{center}
\end{figure}

The dedicated {\it BiPo} detector (see Fig.~\ref{snemo_bipo}) is being developed with the aim of measuring $^{208}$Tl/$^{214}$Bi activities in thin foils
(12~m$^2$, 5 kg) at the level required, with reasonable exposure,
by tagging the {\it Bi}smuth-{\it Po}lonium chain signature: an electron followed by a 
delayed alpha particle~\cite{BiPo}. Three different BiPo prototypes have been developed and tested
in the Underground Laboratory of Modane (LSM,
France)~\footnote{http://www-lsm.in2p3.fr/}.
The background level A($^{208}$Tl)$=$1.5~$\mu$Bq/kg  has been reached after more than one year of measurement with the most successful BiPo-I prototype (0.8~m$^2$ of detecting surface). Extrapolating this level of background the currently in constuction BiPo-III detector (12 m$^2$ of active surface area) will be able to check the radiopurity of the SuperNEMO selenium foils with the required sensitivity with a six month exposure.

\subsection{Tracker}

\begin{figure}[hbt]
\begin{center}
\psfig{figure=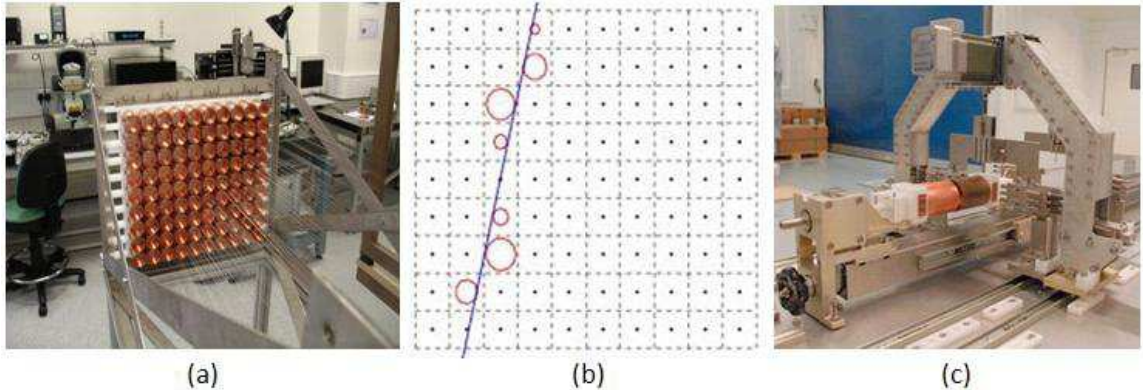,width=0.95\linewidth}
\caption{The 90-cell tracker prototype (a) and reconstructed muon track in it (b); wiring robot (c).}
\label{snemo_tracker}
\end{center}
\end{figure}

Improvement of performance in tagging charged particles (e$^\pm$ and
$\alpha$) in the tracking chamber implies optimal choice of
construction materials, sizes of wires and cell, cell layout
(topology) design permitting automated wiring, working gas mixture,
and readout. The basic cell design has been developed and verified with 
a 90-cell prototype with $\varnothing$4.4~cm x 4~m Geiger cells (see Fig.~\ref{snemo_tracker}-a,b). 
The required performance has been demonstrated on muon data: 0.7~mm transverse and 1~cm longitudinal resolution with $>98$\% cell efficiency.

As an industrial scale is required for assembling the SuperNEMO tracker ($\sim$500,000
wires must be processed), a dedicated wiring robot is being developed
for the mass production of drift cells (see Fig.~\ref{snemo_tracker}-c).

\subsection{Demonstrator}
The first SuperNEMO module, called the demonstrator (see Fig.~\ref{snemo_design_demo}-b,c), will be the first step from R\&D to construction with the aims: 
i) to demonstrate the feasibility of large scale mass production;
ii) to measure backgrounds especially from radon emanation;
iii) to finalize detector design.

Also, the demonstrator will be able to produce a meaningful physics result. With 0.3 expected background events in the 2.8 - 3.2 MeV energy region with 7~kg of $^{82}$Se in 2 years, it is expected to reach a sensitivity $T_{1/2}^{0\nu\beta\beta} = 6.5 \cdot 10^{24}$ y (90\% CL) in 2015, which corresponds to $\sim$~4 ``golden events'' if the $0\nu\beta\beta$-evidence claim is correct~\cite{KK}.

\subsection{Miscellaneous}

\noindent {\it Simulations.} The SuperNemo SoftWare (SNSW)
package has been developed and extensive simulations have been done in order to optimize the SuperNEMO design. They have proved that the SuperNEMO target sensitivity is 
reachable with the target parameters given in Table~\ref{parameters}.

\begin{wrapfigure}{r}{7.3cm}
\begin{center}
\psfig{figure=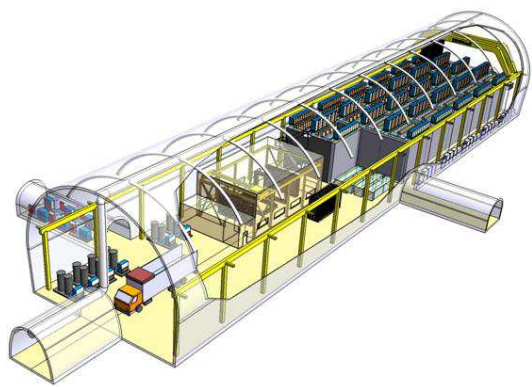,width=7.cm}
\caption{The SuperNEMO housing plan in the Hall~A of the new LSM.}
\label{lsm}
\end{center}
\end{wrapfigure}

\noindent {\it Location.} The SuperNEMO will be located in a new cavern at LSM. One should emphasize that the SuperNEMO is one of the main projects to be hosted in the new LSM laboratory (Hall~A), which is expected to be available in 2013 (see Fig.~\ref{lsm}). If construction of a new hall of LSM will be delayed then the demonstrator will replace the NEMO-3 detector.

The BiPo-III will be hosted in the Canfranc Underground
Laboratory~\footnote{http://ezpc00.unizar.es/lsc/index2.html}.

\noindent {\it Schedule.} The current SuperNEMO plans are the following:
i) the demonstrator construction --- 2010-2012;
ii) the demonstrator physics run start-up --- 2013;
iii) full detector construction start-up --- 2014;
iv) target sensitivity will be reached in 2019.

The BiPo-III detector construction and start is planned during 2010-2011.

\section {Conclusion}

$0\nu\beta\beta$ studies have a potential of discovery to reveal
new fundamental properties of the neutrino  in particular and nature in
general. Several experiments with different techniques are required to
confirm definitely any possible signal observation.

Based on the successful experience of the NEMO detectors, the extensive and intensive SuperNEMO R\&D programme is finishing with construction of the demonstrator started in 2010. In terms of sensitivity and time scale, SuperNEMO is competitive with other world-best $0\nu\beta\beta$-projects (e.g., see review~\cite{bb-review}); the unique technique of the SuperNEMO detector could provide the possibility to
study the origin of $0\nu\beta\beta$-decay in the case of its
discovery.

We acknowledge support by the Grants Agencies of France, the Czech Republic, RFBR (Russia), STFC (UK), MICINN (Spain), NSF, DOE, and DOD (USA).

\end{document}